\documentclass[reprint,amsmath,amssymb,aps,pra,superscriptaddress,nofootinbib,floatfix]{revtex4-2}
\usepackage[utf8]{inputenc}
\usepackage[english]{babel}
\usepackage{hyperref}
\usepackage{bm}
\usepackage{hyperref}
\usepackage{physics}
\usepackage{tabularx}
\usepackage[page]{appendix}
\usepackage{nicefrac}
\usepackage[per-mode=power, 
            exponent-product = \times, 
            inter-unit-product = \ensuremath{{\cdot}}, 
            tight-spacing = true,
            number-unit-product=~,
            parse-numbers=false,]{siunitx}
\AtBeginDocument{\RenewCommandCopy\qty\SI}
\usepackage[dvipsnames]{xcolor}
\usepackage{booktabs}
\usepackage{float}
\usepackage{graphicx}
\usepackage[capitalise]{cleveref}
\usepackage[version=4]{mhchem}
\usepackage{tabularx}
\usepackage{ulem}
\usepackage{enumitem}
\usepackage[
    protrusion=true,
    activate={true,nocompatibility},
    final,
    tracking=true,
    kerning=true,
    spacing=true,
    spacing=nonfrench,
    factor=1000]{microtype}
\hypersetup{
 colorlinks=true, 
 linkcolor=blue, 
 citecolor=blue,        
 filecolor=blue,      
 urlcolor=blue}
\usepackage{titlesec}
\usepackage{caption}
\titleformat{\section}[runin]{\normalfont\bfseries}{\thesection}{10pt}{}[~---~~]
\titlespacing*{\section}{0pt}{0pt}{0pt}
\renewcommand{\textcite}[1]{\citenum{#1}}
\newcolumntype{Y}{>{\centering\arraybackslash}X}
\Crefformat{appendix}{Supplementary Information S.#2#1#3}
\usepackage{comment}

\begin{document}
                                            
\author{Gr\'egory~Moille}
\thanks{These authors contributed equally to this work.}
\affiliation{Microsystems and Nanotechnology Division, National Institute of Standards and Technology,
Gaithersburg, USA}
\affiliation{Joint Quantum Institute, NIST/University of Maryland, College Park, USA}
\author{Pradyoth~Shandilya}
\thanks{These authors contributed equally to this work.}
\affiliation{University of Maryland, Baltimore County, Baltimore, MD, USA}
\author{Jordan~Stone}
\affiliation{Microsystems and Nanotechnology Division, National Institute of Standards and Technology,
Gaithersburg, USA}
\author{Curtis~Menyuk}
\affiliation{University of Maryland, Baltimore County, Baltimore, MD, USA}
\author{Kartik~Srinivasan}
\email{kartik.srinivasan@nist.gov}
\affiliation{Microsystems and Nanotechnology Division, National Institute of Standards and Technology,
Gaithersburg, USA}
\affiliation{Joint Quantum Institute, NIST/University of Maryland, College Park, USA}

\date{\today}
\def\mytitle{All-Optical Noise Quenching of An Integrated Frequency Comb}
\title{\mytitle}
                                                            
\begin{abstract} 
\noindent %
\noindent Integrated frequency combs promise transformation of lab-based metrology into disruptive real-world applications, particularly with octave-spanning microcombs enabling self-referenced optical synthesis and clock implementations. However, the integrated resonators that support microcombs suffer from thermal fluctuations, limiting microcomb use outside laboratories due to the need for bulky feedback systems. Kerr-induced synchronization (KIS) offers a solution by eliminating such electronic servo control through all-optical locking. %
Here, we show how KIS profoundly alters the noise characteristics of soliton microcombs and enables small device footprint to be compatible with low-noise operation. The phase-locking between the dissipative Kerr soliton (DKS) and the injected reference reduces the tooth-to-tooth pump noise propagation -- enabling easier carrier-envelope offset stabilization and uniform spectral performance -- while also damping the impact of intra-cavity fluctuations on the DKS, such as thermo-refractive noise (TRN). %
Our theoretical and experimental results show that KIS modifies the comb noise elastic-tape model, maintaining comb tooth linewidths comparable to the pump lasers', unlike single-pumped systems where linewidths increase by-orders-of magnitude from the comb center to its edges. %
Additionally, KIS quenches intrinsic noise sources at the soliton decay rate, regardless of laser coherence or microring thermo-refractive correlations. %
Using these findings, we demonstrate an octave-spanning microcomb operating below TRN limits, using both free-running lasers and lasers correlated via comb self-referencing, with performance limited only by laser frequency noise.

\end{abstract}
\maketitle

\begin{figure*}[t]
    \centering
    \includegraphics{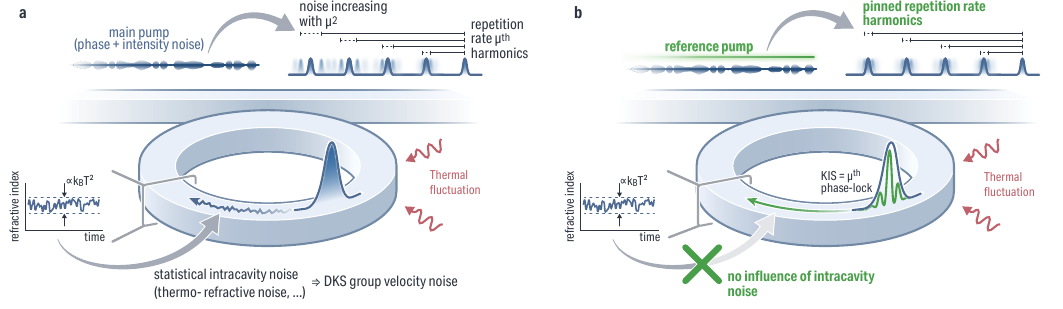}
    \caption{\label{fig:1} \textbf{DKS thermo-refractive noise and its circumvention through Kerr-induced synchronization (KIS).}
    \textbf{(a)} Single pump scenario. The statistical fluctuation of the cavity temperature modifies the cavity material refractive index, which in turn impacts the group velocity of the cavity soliton and hence the repetition rate of the generated frequency comb. In addition, the frequency noise of the pump is transduced onto the cavity soliton group velocity, resulting in noise that increases quadratically between one pulse and another separated by harmonics of the repetition period. In the frequency domain, this corresponds to increasing noise as a function of comb tooth mode number $\mu$ with respect to the pump, which propagates down to the carrier envelope offset (CEO).
    \textbf{(b)} In Kerr-induced synchronization, a reference is sent to the microcavity where the DKS lives, resulting in a phase locking of the cavity soliton to the reference phase velocity. Since the soliton repetition rate is now determined by external parameters that are the two pump frequencies, intrinsic noise sources such as TRN are bypassed and no longer affect the DKS repetition rate. In addition, the microcomb is now dual-pinned at the comb tooth closest to the reference and at the main pump, resulting in noise propagation onto the comb teeth that is damped relative to the single pump case, directly impacting the linewidth of the CEO.
    }
\end{figure*}

Optical frequency combs are essential components in the metrological toolbox because they provide a coherent optical-to-microwave frequency link~\cite{DiddamsScience2020a}, enabling accurate optical frequency measurement\cite{HollbergIEEEJ.QuantumElectron.2001, CundiffRev.Mod.Phys.2003, HallRev.Mod.Phys.2006,GiuntaIEEEPhotonicsTechnol.Lett.2019}, which has found application in optical frequency synthesis~\cite{HolzwarthPhys.Rev.Lett.2000}, spectroscopy~\cite{PicqueNaturePhoton2019a}, microwave signal generation~\cite{FortierNaturePhoton2011}, and ranging~\cite{CaldwellNature2022a}, among others critical applications~\cite{BjorkScience2016, MetcalfOpticaOPTICA2019a, ThorpeOpt.ExpressOE2008}. To further lower the size, weight, and power consumption (SWaP) of combs for deployable applications, on-chip microcombs can be created by taking advantage of nonlinear photonic devices. Leveraging the third-order nonlinearity ($\chi^{(3)}$), integrated microring resonators can be designed to support dissipative Kerr soliton (DKS) states that exist by doubly balancing the loss/gain and dispersion/nonlinear phase shift of the system~\cite{LeoNaturePhoton2010,HerrNaturePhoton2014}, which once periodically extracted create a uniformly spaced pulsed train, and hence an on-chip frequency comb. 
While many applications of table-top frequency combs have been reproduced at the chip-scale thanks to DKSs~\cite{DuttSci.Adv.2018, KudelinNature2024, NewmanOptica2019,DrakePhys.Rev.X2019, RiemensbergerNature2020, SpencerNature2018}, these integrated microcombs fall short in their repetition rate noise performance - primarily due to their poor ability to manage the cavity thermorefractive noise (TRN)~\cite{DrakeNat.Photonics2020, HuangPhys.Rev.A2019, LeiNatCommun2022, LimNatCommun2017}, which increases inversely with the resonator size~\cite{PanuskiPhys.Rev.X2020}. Since TRN locally modifies the material refractive index, the cavity soliton experiences transduction of this noise onto both its group and phase velocity, which respectively impact the repetition rate and the carrier envelope offset (CEO) of the microcomb [\cref{fig:1}a]. This noise transduction is particularly consequential for octave-spanning microcombs, where CEO detection and stabilization is a focus, and where SWaP considerations favor a high repetition rate to maximize individual comb tooth power within an octave, resulting in the use of a small microring resonator~\cite{BraschLightSciAppl2017a,SpencerNature2018, WuOpticaOPTICA2023,MoilleNature2023}. %
Mitigation of TRN is possible, for instance through judicious use of a cooler-laser that can squeeze the resonator temperature fluctuations under low cooler power~\cite{SunPhys.Rev.A2017}, leading to an improvement in the repetition rate noise by about an order of magnitude~\cite{DrakeNat.Photonics2020}. %
Other promising methods such as dispersion engineering to tailor the recoil and counterbalance TRN have been proposed~\cite{StonePhys.Rev.Lett.2020}, and could potentially fully quench it. Yet, such engineering is challenging for octave-spanning combs since the dispersion would need to simultaneously support the desired microcomb bandwidth and the required asymmetry for TRN suppression. A final solution is to strongly suppress TRN by working at cryogenic temperature, which reduces the thermorefractive coefficient by more than two orders of magnitude~\cite{ElshaariIEEEPhotonicsJ.2016a}, and has been demonstrated to enable direct and adiabatic DKS generation~\cite{MoillePhys.Rev.Appl.2019}. Though very effective at removing TRN and other vibrational-based noise sources (\textit{e.g.,} Raman scattering), the significant infrastructure requirements are incompatible fieldable applications.
Beyond repetition rate noise, the individual comb tooth linewidths, as well as the carrier envelope offset tone linewidth, which are directly related to the noise of the main-pump laser generating the DKS through a so-called elastic-taped model~\cite{TelleApplPhysB2002}, can also be a limitation. Here, the frequency noise power spectral density (PSD) of the pump laser propagates quadratically with the comb tooth mode number -- which in the temporal domain corresponds to a quadratic increase of noise between pulses separated by harmonics of the repetition period [\cref{fig:1}a] -- and broadens the comb teeth beyond the linewidth of the pump laser. For octave-spanning combs, where comb teeth that are hundreds of modes away from the pump are used (\textit{e.g.,} in self-referencing), this can be a significant limitation.\\ 
\indent Recently, a new method to stabilize the DKS has been proposed, called Kerr-induced synchronization (KIS)~\cite{MoilleNature2023}. KIS relies on the phase locking of the intra-cavity soliton to an external reference laser injected in the same resonator where the soliton circulates, which locks their respective CEOs [\cref{fig:1}b]. This method has been demonstrated to be highly versatile as any comb line can be synchronized~\cite{WildiAPLPhotonics2023}, and multi-color synchronization that stabilizes the common repetition rate can also be achieved~\cite{Moille2024_colorKIS}. It allows for a passive and all-optical stabilization of the microcomb to an optical frequency reference, providing long-term stability for clockwork applications~\cite{MoilleNature2023} with optical frequency division of the pumps' noise onto the repetition rate~\cite{WildiAPLPhotonics2023, Moille2024_colorKIS, Sun2024}. %
However, the limit of KIS on the noise performance of the microcomb has yet to be determined, and only the impact of the amplitude of either the reference or comb tooth power on the KIS dynamics has been explored. Mathematically, the injection of another laser into the DKS's resonator allows for the existence of a new attractor, regardless of its coherence or TRN anti-correlation with the main pump driving the DKS, which is reached when entering the KIS regime. Since attractors in nonlinear systems only occur when they are dissipative, not only does the amplitude of the energy exchange matter -- like in previous studies -- but so does the characteristic time scale of the energy exchange, which, in our case, is the photon lifetime.

Here, we theoretically analyze and experimentally demonstrate the profound improvements that KIS enables on both the repetition rate noise and individual comb tooth noise in DKS microcombs. We show that the addition of a second intractivity field through the reference critically modifies the elastic-tape model describing the increase of the comb tooth linewidth away from the main pump such that under KIS, all comb tooth linewidths remain within the same order of magnitude, across a span of more than 200 comb lines, while decreasing the free running CEO tone linewidth by two orders of magnitude compared to the singled-pump microcomb. In addition, we demonstrate theoretically and experimentally that synchronization through the reference laser changes the soliton's dynamics such that any internal cavity noise, and in particular TRN, is damped at a rate proportional to the photon lifetime. This behavior, which is expected for dissipative system attractors, essentially bypasses current limitations on microcomb noise performance [\cref{fig:1}b] and enables any cavity size to perform at the same noise level as a large resonator. We experimentally verify our numerical and theoretical analyses using an octave-spanning integrated microcomb with a volume of \qty{\approx80}{\um\cubed}, for which we obtain lower repetition rate noise than the TRN-limited value while using free-running and CEO-locked pumps.

\begin{figure*}[t]
    \centering
    \includegraphics[width = \textwidth]{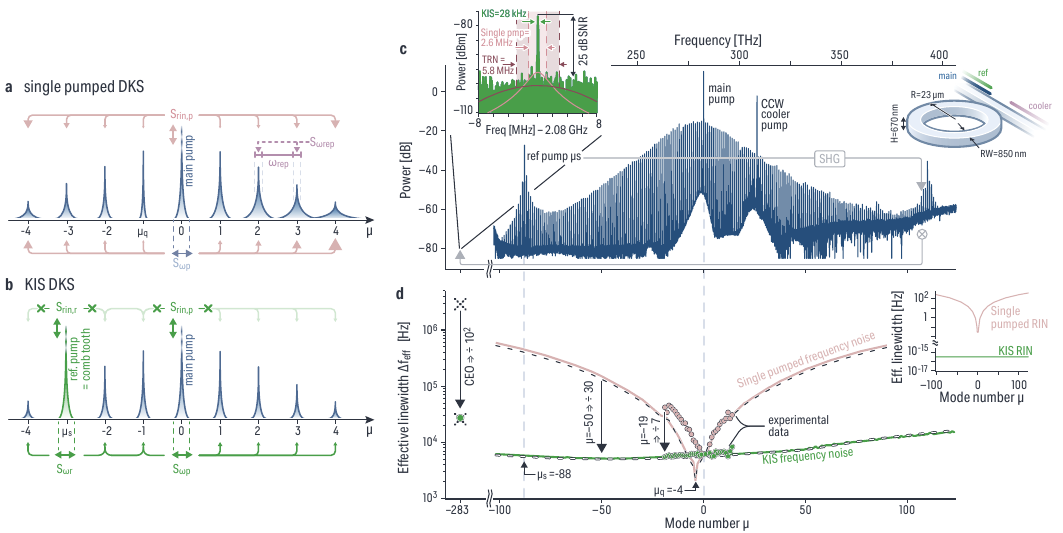}
    \caption{\label{fig:2}%
    \textbf{Impact of KIS on the individual comb teeth linewidths -- }%
    \textbf{(a)} Schematic of noise propagation of the single-pumped microcomb following the elastic-tape model. The frequency noise of the pump $S_\mathrm{\omega,p}$ cascades with a $\mu^2$ scaling relative to a quiet point $\mu_{q}$ that is defined by intrinsic properties of the microring material and dispersion. In addition, the relative-intensity noise of the pump $S_\mathrm{{\omega}rin}$ also propagates onto the frequency noise of the comb with a quadratic scaling $\mu^2$. %
    \textbf{(b)} In KIS, dual-pinning of the comb through the introduction of the reference now prevents any RIN transduction onto the comb teeth frequency noise. While the elastic-tape model for noise propagation still exhibits a quadratic trend $\mu^2$, the dual-pinning greatly limits the propagation of each pumps' noise,  $S_\mathrm{\omega,p}$ and $S_\mathrm{\omega,r}$, onto the comb teeth. %
    \textbf{(c)} Optical spectrum of the experimental microcomb obtained with a main pump at \qty{\omega_\mathrm{pmp}/2\pi\approx 282.4}{\THz} (\qty{1061}{\nm}) and the reference at \qty{\omega_\mathrm{ref}/2\pi\approx 194.5}{\THz} (\qty{1540.1}{\nm}), from a \ce{Si3N4} microring resonator with dimensions \qty{R=23}{\um}, \qty{RW=850}{\nm}, and \qty{H=670}{\nm} (inset). We use a cooler pump at \qty{\omega_\mathrm{cool}/2\pi \approx 308.7}{\THz} (\qty{971.8}{\nm}) for DKS access. By doubling the reference (which is now a comb tooth) and beating it against a comb tooth, we retrieve the carrier envelope offset (inset) with 25~dB SNR (recording bandwidth of \qty{20}{MHz}). %
    \textbf{(d)} Effective linewidths of the individual comb teeth from analytical prediction (dashed lines), LLE simulation (solid lines), and experimental data (circles) with free running laser pump(s). In the single-pump case, the pump frequency noise propagates quadratically with the comb tooth mode number $\mu$, exhibiting a quiet point at $\mu=-4$. In contrast, in KIS noise propagation is damped, as the comb tooth frequency noise must match that of each pump at $\mu = 0$ and $\mu_s=-87$, due to the double pinning of the comb by the two lasers. The experimental data matches well with the KIS model, while a discrepancy is observed for the single pump that is mostly related to thermo-refractive noise that is not yet included in the model. One can expand the elastic tape model down to mode $\mu=-283$, which correspond to the CEO, showcasing a two order-of-magnitude linewidth reduction between KIS and the expected single-pump case. The experimental CEO in KIS agrees with the theoretical model. Importantly, assuming the same beat power, the beat without KIS would be just above the noise level (pink curve in inset of c) preventing any locking. When accounting for TRN as in ref~\cite{DrakeNat.Photonics2020}, the CEO could not be detected (dark red in inset of c). The inset shows the LLE simulation with the pump(s) relative intensity noise (RIN) only, which impacts the comb teeth linewidths in the single pump case (pink), while is totally suppressed in the KIS case (green). Propagation of RIN is suppressed in KIS since the soliton group velocity is insensitive to pump power fluctuations as the repetition rate is pinned by the frequencies of the two pumps.  %
    }
\end{figure*}

\section*{Individual comb tooth linewidth reduction in KIS}%
First, we will discuss the optical linewidth of the individual teeth forming the comb, which is critical in applications such as optical frequency synthesis and spectroscopy, where having narrow individual comb teeth is essential. As discussed previously, we will focus on free-running operation, where two main noise sources from the DKS pump laser can be considered: frequency noise and residual intensity noise (RIN); laser shot noise will be neglected since the frequency noise will be orders of magnitude larger when not locked to a stable reference cavity. Other intrinsic noise of the microring resonator, such as TRN, will be discussed later in the paper.\\
\indent In the single-pumped DKS, noise propagation from the pump, either from its frequency noise or RIN, to the microcomb teeth follows the well-known elastic-tape model~\cite{TelleApplPhysB2002}. Interestingly, the frequency noise must account for soliton-recoil~\cite{LeiNatCommun2022}, which can appear from the Raman self-frequency shift~\cite{MilianPhys.Rev.A2015, YiOpticaOPTICA2015}, and/or dispersive wave-induced rebalancing where the center of mass of the DKS must be null~\cite{CherenkovPhys.Rev.A2017}. Here, the Raman effect is neglected since its impact is marginal in \ce{Si3N4} compared to other materials~\cite{KarpovPhys.Rev.Lett.2016}, and we assume that the recoil mostly comes from the imbalance of the DW powers creating a shift of center of mass. The PSD of a comb tooth $\mu$, referenced to the main pump ($\mu=0$), can be written as $S^{1p}(\mu,f) = S^{1p}_\mathrm{{\omega}rin} (\mu,f) + S^{1p}_\omega (\mu,f)$, with $S^{1p}_\mathrm{{\omega}rin}(\mu, f)  = S_{{\omega}rin,p}(f)\mu^2$, where $S_{{\omega}rin,p}$ already accounts for the power-to-frequency noise transduction from the pump and the frequency noise PSD defined as described in ref~\citenum{LeiNatCommun2022} as: 
    \begin{equation}
        S^{1p}_\omega (\mu, f)= S_\mathrm{\omega,p}(f)\left(1 - \frac{\partial \omega_\mathrm{rep}}{\partial \omega_p}\mu\right)^2
    \end{equation} 
Here, $S_\mathrm{\omega,p}$ is the frequency noise PSD of the main pump, $\omega_\mathrm{rep}$ is the repetition rate, and $\omega_p$ is the main pump frequency. Therefore, the main pump noise cascades onto the comb teeth at a quadratic rate [\cref{fig:2}a]. The superscript \textit{1p} refers to the single-pump DKS operation, which will be compared against the KIS regime. It thus becomes obvious that, apart from a quiet comb tooth at $\mu_q = \left(\nicefrac{\partial \omega_\mathrm{rep}}{\partial \omega_p}\right)^{-1}$, the noise propagation makes the individual comb teeth broader than the pump itself, in particular for teeth far from the pump. As such comb teeth are of critical importance in octave-spanning combs, this noise propagation is a hindrance on performance.

Since KIS involves phase locking of the frequency components at the reference mode $\mu_s$, resulting in the capture of the comb tooth closest to the reference pump laser, the noise propagation from the pump lasers will be largely different from the single pump regime [\cref{fig:2}b]. First, the microcomb is now fully defined by the frequency of both lasers, regardless of their power. Thus, as long as the cavity soliton remains within the KIS bandwidth, which is dependent on the DKS $\mu_s$ modal component and  reference intracavity energies~\cite{MoilleNature2023, Moille2024_ACKIS} and is $\approx$ \qty{1}{\GHz} here, the impact of RIN from both pumps on the individual comb teeth will be entirely suppressed, such that $S^{kis}_\mathrm{{\omega}rin} (\mu) = 0$ [\cref{fig:2}d-inset], where  the \textit{kis} superscript refers to the KIS regime throughout the manuscript. Hence, in our approximation to dismiss for the moment the microring intrinsic noise, the only contribution to the individual comb tooth noise PSD $S^{kis}(\mu)$ comes from the frequency noise cascading of both pumps. Since the captured comb tooth and the reference pump became indistinguishable in KIS, the comb tooth frequency noise is now pinned at each of the two pumps. In addition, the repetition rate noise must be the same for any two adjacent comb teeth considered. Hence, the noise propagation must also follow an elastic-tape model, this time adjusted for the dual-pinning such that:
    \begin{equation}
        S^{kis}_\omega (\mu, f)= S_\mathrm{\omega,p}(f) \left(1- \frac{\mu}{\mu_s}\right)^2  + S_\mathrm{\omega ,r}(f)\left(-\frac{\mu}{\mu_s}\right)^2
    \end{equation}
    with $S_\mathrm{\omega,r}$ the frequency noise PSD of the reference pump and $\mu_s$ the comb tooth order undergoing KIS. It is interesting to compare the frequency noise in both regimes. Assuming the same frequency noise for both pump lasers $S_{\omega, p} = S_{\omega, r}$, and that $S_\mathrm{{\omega}rin}(\mu) \ll S_\omega(\mu)$, which is obviously true in the KIS case and has been demonstrated in the single pump regime~\cite{LeiNatCommun2022}, one will observe $S^{1p}_\omega<S^{kis}_\omega$ for the modes $\mu = [\nicefrac{2 \mu_{s} \mu_\mathrm{q} \left(\mu_{s} - \mu_\mathrm{q}\right)}{\mu_{s}^{2} - 2 \mu_\mathrm{q}^{2}}\; ; \;0]$. One of the main advantage of KIS is the possibility to dual-pin the microcomb's widely-separated modes~\cite{MoilleNature2023, WildiAPLPhotonics2023}, while the recoil responsible for $\mu_q$ is usually close to the main pump. Hence, assuming $\mu_s \gg \mu_\mathrm{q}$, the single pump comb teeth PSD will only be better than the KIS one for a range $\mu = [2\mu_q; 0]$, which by definition is narrow since $\mu_q$ is small. In contrast, the KIS microcomb will exhibit better single comb line frequency noise PSD performance across a broadband spectrum. Interestingly, in KIS when $S_{\omega,p} \approx S_{\omega,r}$, a new quiet point at $\mu_q = \mu_s S_{\omega,r} / (S_{\omega,p} + S_{\omega,r})$ will be present with a PSD reduction of twice from the pump's PSD average. However, in the case of $S_{\omega,r} \gg S_{\omega,p}$ ($S_{\omega,r} \ll S_{\omega,p}$) the lowest comb tooth noise will be at the main pump $\mu_q =0$ (reference pump $\mu_q= \mu_s$).

To demonstrate such performance improvement through KIS, we use an integrated microring resonator made of \qty{H=670}{\nm} thick \ce{Si3N4} embedded in \ce{SiO2}, with a ring radius of \qty{R=23}{\um} and a ring width of \qty{RW=850}{\nm}. This design provides anomalous dispersion around \qty{283}{\THz} (\qty{1060}{\nm}) with higher-order dispersion providing nearly harmonic dispersive waves (DWs) at \qty{194}{\THz} and \qty{388}{\THz}, corresponding to $\mu=-88$ and $\mu=110$ respectively [\cref{fig:2}c]. We measure the individual comb lines from \qtyrange{\mu=-19}{\mu = 14}{} using a high-rejection and narrowband optical filter and an optical frequency discriminator (see Supplementary Information S.6 for details), while both main and reference pumps remain free running with an effective linewidth \qty{\Delta f_\mathrm{eff} \approx 5.8}{\kHz} determined from the laser PSD measurement following the definition~\cite{HjelmeIEEEJ.QuantumElectron.1991} $\int_{\Delta f_\mathrm{eff}}^{+\infty} \nicefrac{S_{\omega}(f)}{f^2}\;\mathrm{d}f = \nicefrac{1}{\pi}$. In the KIS regime, the measured comb teeth effective linewidths remain in the \qtyrange{5}{7}{\kHz} range, obtained from their frequency noise spectra in a similar fashion as the pumps. The experimental data closely matches Lugiato-Lefever equation (LLE) simulations accounting for both pumps~\cite{TaheriEur.Phys.J.D2017}, which in addition matches the dual-pinned elastic-taped model introduced above. Here, the experimentally measured frequency noise PSD for each pump is input into the LLE simulation. Such close agreement between the model and the experimental data enables us to predict a comb tooth linewidth that will not exceed \qty{\Delta f_\mathrm{eff} = 15}{\kHz} for modes as far as $\mu = +120$. 
In sharp opposition, the predicted noise of the single pump case, which as expected from previous work~\cite{TelleApplPhysB2002,LeiNatCommun2022} matches between the LLE and analytical model assuming a quiet comb line at $\mu_p = -4$ obtained from the maximum of the $\mathrm{sech}^2$ envelope of the experimental microcomb, yields an effective linewidth above \qty{50}{kHz} ($10\times$ that of the main pump) for modes below $\mu=-28$ or above $\mu =22$, and as high as \qty{500}{\kHz} for modes close to the DWs. Experimentally, we confirmed such a trend, where $\Delta f_\mathrm{eff}$ increases with separation from the main pump, with \qty{\Delta f_\mathrm{eff} \approx 40}{\kHz} for $\mu=-19$. However, we do not observe a quiet mode in the measurements, and instead all comb lines exhibit higher noise than the pump. This discrepancy, which has been observed previously~\cite{LeiNatCommun2022}, is mostly related to TRN that is not yet accounted for in the model. The lack of impact of its absence in modeling the KIS results already hints at its quenching in that regime, while it is the predominant noise source in the single pump case.\\
\indent Importantly, the modification of the elastic tape model enables a quantitative comparison of the carrier envelope offset (CEO) linewidth. Under KIS, the CEO linewidth is reduced by approximately two orders-of-magnitude compared to the single-pump case, from \qty{2.6}{\MHz} to \qty{27.1}{\kHz}, assuming only the frequency noise of the pump. We experimentally confirm this through $f-2f$ interferometry between the doubled reference laser (now a comb tooth) and its nearest comb tooth, measuring a KIS CEO linewidth of \qty{28}{\kHz} [\cref{fig:2}c inset]. This drastic reduction in the CEO linewidth enables a larger signal-to-noise ratio and easier CEO locking, as we will discuss later. Additionally, we can estimate the CEO linewidth induced by thermo-refractive noise (TRN) to be approximately 5.8 MHz, using the system parameters presented in Supplementary Information Table S.II and following the analysis in~\cite{DrakeNat.Photonics2020}. This also suggests that TRN plays a significant role in the single-pump case, which is largely mitigated in the KIS case.  %

Additionally, LLE simulations confirm our intuition that pump RIN has been completely removed from influencing the effective linewidths of the comb teeth in the KIS regime [\cref{fig:2}d inset]. In contrast, pump RIN is a non-negligible contribution in the single-pump case, although it remains orders-of-magnitude lower than the pump laser frequency noise. %
Such modification of the elastic tape model for comb tooth noise could also provide valuable insights into other systems where similar injection locking of a comb tooth occurs. Recently, it has been demonstrated that the recycling of a comb tooth for self-injection locking back into the comb~\cite{LeiOpticaOPTICA2024} results in a reduction in the comb teeth linewidths over a large modal span, akin to what is presented here. A similar analysis and modification of the elastic tape model, like the one we proposed, could be used to predict the comb teeth linewidths in such a system.

\vspace{1em}
\section*{Theoretical study of intrinsic noise suppression in Kerr-induced synchronization}%

\begin{figure*}[t]
    \centering
    \includegraphics{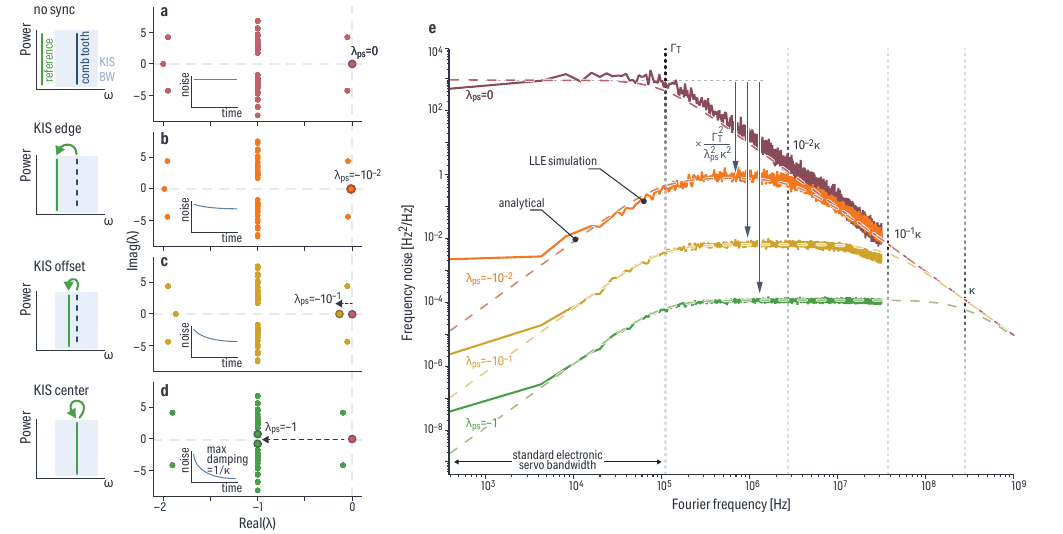}
    \caption{\label{fig:3}
    \textbf{Theoretical study of intrinsic noise suppression in KIS -- }%
    \textbf{(a)} Eigenvalue spectrum of the linearized operator (dynamical spectrum) in the unsynchronized case. The highlighted eigenvalue (large red circle) is the position-shifting eigenvalue $\lambda_{ps}$, which in this case is null and corresponds to perturbations, like thermo-refractive noise (TRN) fluctuation of the DKS group velocity, persisting (inset). 
    A similar study as in (a), with the reference at \textbf{(b)} the edge of the KIS window corresponding to $\lambda_\mathrm{ps} = -10^{-2}$, \textbf{(c)} offset from the center of the KIS window at $\lambda_\mathrm{ps} = -10^{-1}$, and \textbf{(d)} at the center of the KIS window, that is, in the original location of the soliton comb tooth, corresponding to the minimum value of $\lambda_{ps}=-1$. In each case, the noise is damped at a rate of $\lambda_{ps}\kappa$, where $\kappa$ represents the photon loss rate (inset). %
    \textbf{(e)} The power spectral density of the repetition rate noise, $S_\mathrm{rep}$, is numerically simulated using the LLE (solid line) and is in accordance with the analytical analysis presented in~\cref{eq:psd_kis} (dashed lines). In the unsynchronized case where $\lambda_\mathrm{ps}=0$ (red), the noise follows the pattern determined by the TRN, featuring a high-frequency cut-off at $\Gamma_T\ll\kappa$. When $\lambda_\mathrm{ps}<0$, indicating a synchronized case, the noise is damped at a rate of $\lambda_{ps}\kappa$. This leads to a significant reduction of the noise below $\Gamma_T$ and forms a plateau for frequencies between $\Gamma_T$ and $|\lambda_\mathrm{ps}|\kappa$. This plateau exhibits a decrease from the single-pump plateau by a factor of $\Gamma_T^2/\lambda_\mathrm{ps}^2\kappa^2$. For frequencies exceeding the KIS actuation bandwidth $|\lambda_\mathrm{ps}|\kappa$, the noise follows the pattern of the single-pump TRN, which aligns well with the LLE simulation for $\lambda_\mathrm{ps}=-10^{-2}$ (orange). However, for larger values of $|\lambda_\mathrm{ps}|$, the LLE simulation becomes too computationally demanding to resolve at high frequency a cut-off that is analytically confirmed. The maximum damping is achieved at the center of the KIS window ($\lambda_\mathrm{ps}=-1$), with the widest KIS actuator bandwidth at $\kappa$ (green).}%
\end{figure*}

Since we observed good agreement between KIS experimental data and the theoretical dual-pinned elastic tape model that only accounts for pump frequency noise, even with mode separation down to CEO, while the single pump case presents a discrepancy which can likely be attributed to TRN, we now focus on studying the microring resonator's intrinsic noise and how KIS modifies its impact on the cavity soliton. With this in mind, we study noise of the repetition rate $\omega_\mathrm{rep}$ of the comb rather than noise of individual comb lines, since we have shown that regardless of operating in KIS or the single pump regime, noise propagates through an elastic-tape model. We proceed with a linear stability analysis of the LLE for the cavity soliton outside of synchronization (\textit{i.e.}, akin to single pump operation) and in the KIS regime.

Our starting point is the multi-driven LLE (MLLE)~\cite{TaheriEur.Phys.J.D2017,HanssonPhys.Rev.A2014}, normalized to losses (see Supplementary Information S2 for details), which can be written as:

\begin{align}
    \label{eq:MLLE}
    \frac{\partial \psi}{\partial t} &= -(1 + i\alpha_\mathrm{pmp})\psi + i\sum_{\mu}\mathcal{D}(\mu)\tilde{\Psi} \mathrm{e}^{i\mu \theta} + i|\psi|^2\psi \\
    &+ F_\mathrm{pmp} + F_\mathrm{ref} \exp\Bigl[ i\Bigl(\alpha_\mathrm{ref} - \alpha_\mathrm{pmp} + \mathcal{D}(\mu_s)\Bigr)t - i\mu_s\theta\Bigr]\nonumber
\end{align}
where $\psi$ is the normalized intracavity field of the DKS following ref.~\textcite{TaheriEur.Phys.J.D2017} and detailed in the Supplementary Information S2, $\theta$ is the azimuthal coordinate of the ring, $\mu$ is the azimuthal mode number, $\tilde{\Psi}$ is the Fourier transform of $\psi$ from $\theta$ to the $\mu$ domain, $t$ is the normalized time, $\alpha_\mathrm{pmp}$($\alpha_\mathrm{ref}$) is the normalized detuning of the primary (reference) pump, $F_\mathrm{pmp}$ ($F_\mathrm{ref}$) is the intra-cavity power of the primary (reference) pump, $\mathcal{D}(\mu)$ is the normalized integrated dispersion of the cavity as a function of relative mode number $\mu$, and $\mu_s$ is the mode number of the reference pump in KIS. 

In general, \cref{eq:MLLE} admits multi-color soliton solutions. In the KIS regime, the soliton solution does not change from the single pumped one except for a constant drift which is associated with the time-dependence of the reference pump field. Additionally, the reference pump field and the DKS field are stationary with respect to each other in the KIS regime. We can change the frame of reference in \cref{eq:MLLE} to obtain an equation with stationary solutions:
\begin{equation}
    \theta - \frac{\left( \alpha_\mathrm{ref} - \alpha_\mathrm{pmp} + \mathcal{D}(\mu_s) \right)}{\mu_s}t = \theta',~ t = t'.
\end{equation}
This frame of reference rotates with an angular velocity where the reference pump is stationary. Using

\begin{align}
    \frac{\partial}{\partial \theta} &= \frac{\partial}{\partial \theta'}, \\
    \frac{\partial}{\partial t} &= - \frac{\left( \alpha_\mathrm{ref} - \alpha_\mathrm{pmp} + \mathcal{D}(\mu_s) \right)}{\mu_s} \frac{\partial}{\partial \theta'} + \frac{\partial}{\partial t'},
\end{align}
we can rewrite \cref{eq:MLLE} as:
\begin{align}\label{eq:CFR}
    \frac{\partial \psi}{\partial t} &= -(1 + i\alpha_\mathrm{pmp})\psi + i\Sigma_{\mu}\mathcal{D}(\mu)\Psi \mathrm{e}^{i\mu \theta} \nonumber \\&+ \frac{(\alpha_\mathrm{ref} - \alpha_\mathrm{pmp} + \mathcal{D}(\mu_s))}{\mu_s}\frac{\partial \psi}{\partial \theta} + i|\psi|^2\psi \\&+ F_\mathrm{pmp} + F_\mathrm{ref} \exp\left[- i\mu_s\theta\right]. \nonumber
\end{align}

For convenience, we have dropped the primes in \cref{eq:CFR}. The MLLE in the form of \cref{eq:CFR} admits stationary solutions in the KIS regime as the right-hand side is time-independent. Now, the stability of the soliton in the presence of perturbations can be studied using dynamical techniques. For the given set of parameters, we can calculate the stationary solution $\psi_0$ using the Levenberg-Marquardt algorithm such that $\frac{\partial \psi_0}{\partial t} = 0$. Following the procedure described in Ref~\citenum{WangJ.Opt.Soc.Am.B2018a}, we linearize \cref{eq:CFR} around $\psi_0$ to obtain:
\begin{equation}\label{eq:linearized}
    \frac{\partial \Delta\psi}{\partial t} = \mathcal{L}(\psi_0)\Delta\psi,
\end{equation}
where $\Delta\psi$ is a perturbation and $\mathcal{L}(\psi_0)$ is the linearized operator obtained numerically. The eigenvalues of this operator are referred to as the spectrum of the linearized operator or the dynamical spectrum, and they determine the local stability of the system~\cite{MenyukWangNanophotonics2016,QiMenyukOptica2019}.  Solving \cref{eq:linearized}, we obtain the effect of the system on the perturbation after time $\Delta t$:
\begin{equation}
    \Delta\psi(t_0 + \Delta t) = \exp\left[\mathcal{L}(\psi_0) \Delta t\right]\Delta\psi(t_0).
\end{equation}
In general, one can decompose any perturbation into a linear combination of eigenfunctions $v_n$ of $\mathcal{L}(\psi_0)$, with corresponding eigenvalues $\lambda_n$. Therefore,
\begin{align}\label{eq:perturbation}
    \Delta \psi(t_0) &= \sum_n a_n v_n, \nonumber\\
    \Delta \psi(t_0 + \Delta t) &= \sum_n \exp[\lambda_n \Delta t]a_n v_n.
\end{align}
From \cref{eq:perturbation}, it is clear that a perturbation to the DKS grows exponentially if $\Re(\lambda_n) > 0$, persists if $\Re(\lambda_n) = 0$, and damps exponentially if $\Re(\lambda_n) < 0$. In all cases, there is one eigenvalue $\lambda_{ps}$, referred to as the position-shifting eigenvalue, whose eigenfunction corresponds to perturbations of the DKS position inside the cavity. Perturbations of the form of the position-shifting eigenfunction are responsible for fluctuations in the repetition rate of the soliton. For the singly-pumped DKS [\cref{fig:3}a], $\lambda_{ps} = 0$, which implies that perturbations to the soliton position due to noise persist. $\lambda_{ps}$ in the singly-pumped DKS exists at zero because the DKS has a degree of freedom associated with its translational invariance, which is due to the fact that the singly-pumped microresonator system is perfectly symmetric in the
resonator’s azimuthal coordinate. However, in the KIS regime, the soliton is trapped by the two pumps, and results in $\lambda_{ps}<0$ [\cref{fig:3}(b)-(d)], which implies that perturbations to the DKS damp exponentially. In other words, the injection of the reference pump breaks the symmetry in the resonator, meaning that the DKS does not exhibit any translational invariance anymore. When the reference pump's frequency is at the center of the synchronization window, $\lambda_{ps} = -1$ [\cref{fig:3}d], which corresponds to exponential damping of the DKS jitter with the photon lifetime. This is the most negative value that $\lambda_{ps}$ can take as all photons in the cavity take on average a photon lifetime to exit, and therefore sets the fundamental limit for damping of the repetition rate noise, which after accounting for the normalization with respect to the total loss \qty{\kappa=280}{\MHz}, corresponds to exponential damping of the intra-cavity noise by the photon lifetime $\tau_\mathrm{phot}= \nicefrac{1}{\kappa}$.  A physically intuitive picture as to why the photon lifetime is important can be obtained as follows. The repetition rate noise manifests as a perturbation to the soliton's intracavity mode frequencies. In the KIS regime, the soliton begins to converge to an equilibrium from this perturbed state. However, the photons that were a part of the perturbation have to exit the cavity in order for the system to be in equilibrium. Owing to the $Q$ of the cavity, these photons leave the cavity on average in a photon lifetime. Therefore, a lower photon lifetime enables the noise to damp faster. 
To verify the stability analysis conducted, we performed stochastic LLE simulations accounting for intra-cavity TRN, for which the repetition rate can be extracted at every-round trip time and processed to obtain its PSD. These simulations account for the impact of TRN on pump detunings $\alpha_{p,r}$ and the dispersion $\mathcal{D}$~\cite{DrakeNat.Photonics2020, StonePhys.Rev.Lett.2020}, and using either simulated or experimentally determined values for other parameters (see Supplementary Information S.1). In the unsynchronized case [\cref{fig:3}a], the simulated repetition rate noise matches well with the analytical one that can be obtained following ref.~\cite{StonePhys.Rev.Lett.2020}:
\begin{equation}
    \label{eq:psd_reprate_1pmp}
    S_\mathrm{rep}^{1p}(f) = 2\frac{\eta_T^2 k_B T^2}{\rho C V} \frac{\Gamma_T}{(f^2 + \Gamma_T^2)}
\end{equation}

\noindent with $\eta_T = \left .\nicefrac{\partial \omega_{\text{rep}}}{\partial T}\right |_{T_0}$ calculated when not synchronized (details in Supplementary Information S.3), $\rho$ the material density, $C$ its specific heat, and $V$ the modal volume of the fundamental transverse electric mode, $\Gamma_T$ the thermal dissipation rate, $T$ the temperature, $k_B$ the Boltzmann constant, and $f$ the Fourier frequency. Details for the parameter values and the methods by which they were obtained can be found in Supplementary Information S.4. Consistent with the linear stability analysis, the $\omega_\mathrm{rep}$ PSD result from the stochastic LLE exhibits a much lower noise level in the KIS regime [\cref{fig:3}d], which is minimized when the reference is at the center of the KIS bandwidth (\textit{i.e.}, where the comb tooth is originally). In our simulation, the frequency noise of the pumps has been neglected, yielding driving fields that are perfect Dirac delta functions spectrally and hence resulting in the reduction of the PSD at low Fourier frequencies. Interestingly, one could derive an analytical expression for the repetition rate noise in the KIS regime, which can be obtained from the linear stability analysis of the MLLE regime:

\begin{equation}
    \frac{\partial\Delta \omega_\mathrm{rep}}{\partial t} = \kappa \lambda_{ps}\Delta \omega_\mathrm{rep} + \eta_T \frac{\partial \Delta T}{\partial t}
\end{equation}

\noindent with $\kappa = 1/\tau_\mathrm{phot}$ the total loss rate of the cavity and $\tau_\mathrm{phot}$ the photon lifetime, $\lambda_{ps}$ the position-shifting eigenvalue, and $\Delta \omega_\mathrm{rep}$ the repetition rate noise. The PSD of the repetition rate noise can then be derived following Supplementary Information S.1 such that:
\begin{equation}
    \label{eq:psd_kis}
    S_\mathrm{rep}^{kis}(f) = S_\mathrm{rep}^{1p}(f) \times \frac{f^2}{ f^2+\lambda_{ps}^2\kappa^2}
\end{equation}

Outside of synchronization $\lambda_{ps}=0$ yields the same expression as the single pump case~(\cref{eq:psd_reprate_1pmp}) with a typical Lorentzian profile in the Fourier frequency space. %
In the KIS regime, the noise spectrum is then damped by the $\nicefrac{f^2}{(f^2+\lambda_{ps}^2\kappa^2)}$ term, linking KIS to the energy exchange rate of the system, as expected from attractors of nonlinear dissipative systems, and is in good agreement with the simulations [\cref{fig:3}d], albeit a discrepancy at low Fourier frequency due to the finite simulation finite time and averaging. In comparison to the conventional single pump DKS, the PSD profile of $\Delta\omega_\mathrm{rep}$ is dramatically altered, since at low frequency it follows a $\nicefrac{S_0f^2}{\Gamma_T^2\lambda_{ps}^2\kappa^2}$ profile, with  $S_0=2\frac{\partial \omega_\mathrm{rep}}{\partial T}\frac{k_B T^2\Gamma_T}{\rho C V}$. As the thermal dissipation rate is much slower than the photon decay rate ($\Gamma_T\ll \kappa$) in the microring, this hints at the good long term stability of the repetition rate, which will no longer be hindered by incoherent intracavity noise processes such as TRN, in contrast to the single pump case that exhibits a typical plateau at $\nicefrac{S_0}{\Gamma_T^2}$ at low Fourier frequencies.
At high Fourier frequency $f> |\lambda_{ps}|\kappa$, the PSD in KIS will follow the Lorentzian profile imposed by the thermal noise, yet the Fourier frequency at which it happens is $\lambda_{ps}$ dependent since a plateau at $\nicefrac{S_0}{\lambda_{ps}^2\kappa^2}$ will be observed for $f \in [\Gamma_T; |\lambda_\mathrm{ps}|\kappa]$; however, this plateau occurs at a value that is still much lower than the noise in the single pump case, which follows a $\nicefrac{1}{f^2}$ trend, and reduced from the single-pump low-frequency plateau by a factor $\nicefrac{\Gamma_T^2}{\lambda_\mathrm{ps}^2\kappa^2}$. Physically, it is important to note that the statistical fluctuations of the refractive index of the material leading to TRN remain present in the microresonator; however, KIS provides a trapping of the DKS. The dynamics of attractors in non-linear dissipative systems tells us that any perturbations faster than their characteristic energy exchange rate cannot be counteracted. In KIS, the exchange rate being defined by the photon lifetime $\tau_\mathrm{phot} = 1/\kappa$ and the reference laser frequency relative to the KIS bandwidth fixing the zero mode eigenvalues $\lambda_\mathrm{ps}$, any noise faster than $|\lambda_\mathrm{ps}|\kappa$ will not be influenced by the soliton synchronization and will be experienced as in the single-pumped regime. %
The dependence on $\lambda_\mathrm{ps}$ can be understood such that at the edge of the KIS bandwidth, the synchronization is slower since the soliton must adapt its repetition rate within the nonlinear Kerr effect timescale. This results in a small $\lambda_{ps}$ and hence $S_\mathrm{rep}^{kis}$ catches up with $S_\mathrm{rep}^{1p}$ at relatively low frequencies. In contrast, at the center of the KIS bandwidth, synchronization is faster since the reference pump is already at the frequency for which the soliton exists in the single pump case. This results in a larger $|\lambda_{ps}|$ and a higher frequency at which $S_\mathrm{rep}^{kis}$ catches up with $S_\mathrm{rep}^{1p}$. The important aspect of KIS is that the system noise is not limited by material property nor the cavity volume but instead becomes photon-lifetime-limited. Since it limits both the loading time for which the reference can be injected in the microring resonator and the photon dissipation of the previously unsynchronized cavity soliton.  The above results are similar to the so-called ``quantum diffusion limited'' counter-propagative solitons~\cite{BaoNat.Phys.2021}. Here, the theoretical timing jitter limitation of solitons are obtained through a Lagrangian approach~\cite{MatskoOpt.ExpressOE2013}, where an equivalent to KIS occurs between solitons in each direction~\cite{YangNaturePhoton2017}, since both systems obey an analogous Alder equation. This analysis yields a noise limited by the photon lifetime $\nicefrac{1}{\kappa}$. Importantly, here we expand such a quantum diffusion limit to a single soliton, where the synchronization is enabled through entirely controllable external parameters provided by the two laser pumps, instead of using synchronization between two soliton states. We note that while ultra-high $Q$ provides net benefits in terms of reducing the pump power needed to generate Kerr solitons, when it comes to minimizing the impact of intrinsic noise, it is advantageous to reduce $Q$ to have a photon lifetime $\nicefrac{1}{\kappa}$ as short as possible. In this context, \ce{Si3N4} microring resonators present a critical advantage under KIS compared to much larger volume crystalline resonators with ultra-high-$Q$ (\textit{e.g.,} 10$^9$)~\cite{LecaplainNatCommun2016}, since our microring exhibits a decrease in the photon lifetime by about three orders of magnitude while retaining a small enough modal volume to enable relatively low pump power DKS existence, and no longer suffers TRN-related limitations thanks to KIS. %
Importantly, our conclusions provide precision to the current state of understanding of KIS-like systems. Specifically, it has been hypothesized that the synchronized soliton may remain susceptible to TRN~\cite{WildiAPLPhotonics2023}, in a similar fashion to single-pumped DKS. Using this understanding, to obtain a sub-TRN DKS repetition rate noise, the cavity TRN shall be reduced through, for instance, anti-correlated main and reference pumps that counteract the TRN effects on the cavity; such an argument has been brought forward to explain below-TRN synchronized-DKS repetition rate noise~\cite{ZhaoNature2024}. %
In sharp contrast, our theoretical demonstration shows that the soliton adapts its group velocity to compensate for refractive index variations due to TRN, maintaining a fixed repetition rate dictated by the KIS dual-pinning of the comb within the actuation bandwidth allowed by the soliton lifetime. Therefore, even if anti-correlated main and reference pumps reduce the cavity TRN, our results show that this dual-pinning fundamentally changes the dynamics of the DKS and decouples it from the intracavity noise for Fourier frequencies below the inverse of the photon lifetime. Thus, the KIS dual-pinning does not require any common coherence between the pumps for sub-TRN DKS noise, which we aim to also demonstrate experimentally.

\begin{figure*}[t]
  \centering
  \includegraphics[width = \textwidth]{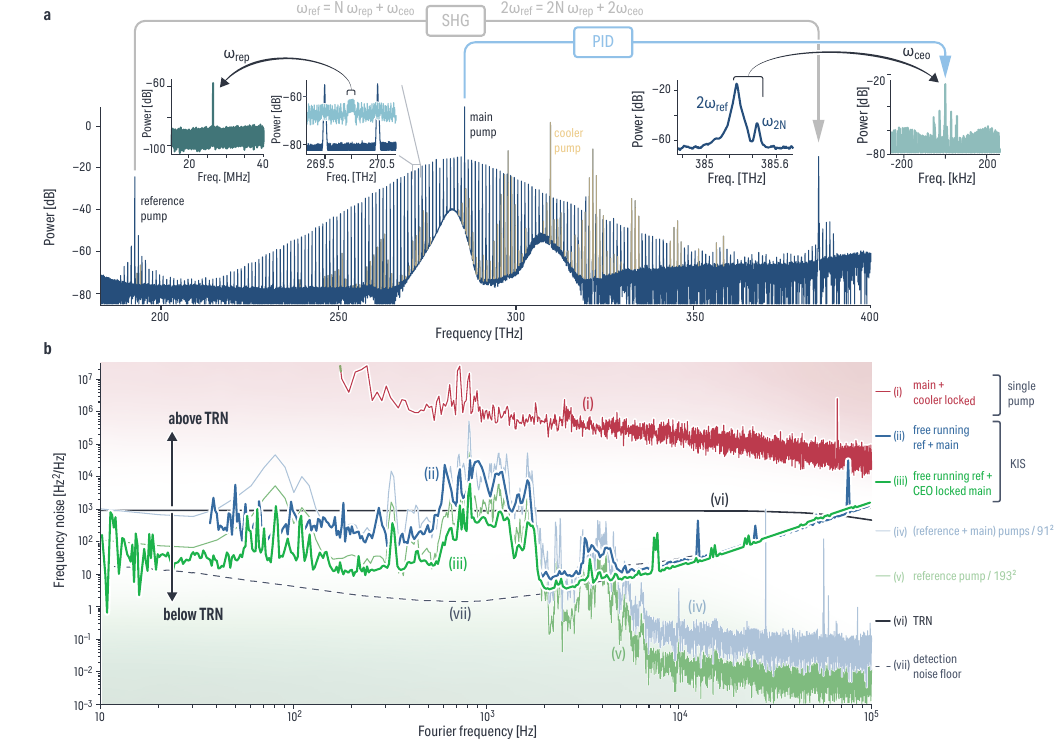}
  \caption{\label{fig:4}%
  \textbf{Themo-refractive noise quenching in an octave-spanning microcomb: free-running and self-referenced -- }%
  \textbf{(a)} Experimental spectrum of the octave-spanning microcomb pumped at \qty{\omega_\mathrm{rep}\approx285.5}{\THz} with the reference pump at \qty{\omega_\mathrm{rep}\approx192.6}{\THz}. A strong cooler pump is used to thermally stabilize the temperature of the resonator which, while being cross-polarized and sent in opposite direction to the main pump, results in some nonlinear interaction (yellow spectrum). The frequency doubled reference is obtained through a second-harmonic generation (SHG) nonlinear crystal and beat against the closet comb tooth at $2f$. We measure the beat note using an electro-optic comb and lock the carrier-envelope offset frequency (CEO) to a \qty{10}{\MHz} Rb frequency standard through feedback to the main pump frequency. The repetition rate is measured around \qty{270}{\THz} with an electro-optic comb that spectrally translates two adjacent microcomb teeth through higher-order sidebands, enabling detection within a \qty{50}{\MHz} bandwidth. 
  \textbf{(b)} Frequency noise obtained for a single pumped microcomb with frequency-locked cooler (i, red), KIS microcomb with free-running main and reference pumps (ii, dark blue) and KIS microcomb with free-running reference pump and locked CEO through feedback to the main pump (iii, green). The frequency noise of the pumps, which is similar for both reference and main pump, is overlaid in thin lines for each KIS case after accounting for the relevant optical frequency division (OFD) factor. When both pumps are free-running, the $OFD=91^2$ corresponds to the comb tooth separation between the pumps (v), and accountin for both pumps not sharing any common noise, is in good agreement with the detected repetition rate noise. In the CEO-locked case, $OFD=193^2$ (vi), which further reduces the repetition rate noise (here the noise on the CEO lock can be neglected and only the free-running reference pump noise is a factor). In both KIS cases, the single-pumped (without cooler) TRN limit (vi, black solid line) is beaten. The detection noise floor (vii, dashed line) is limited by the RF synthesizer used in the electro-optic comb repetition rate measurement apparatus. 
  }
\end{figure*}
\vspace{1em}
\section*{Experimental demonstration using octave-spanning self-referenced comb}
Finally, we experimentally demonstrate that KIS quenches the intrinsic noise of the repetition rate (and thus of the individual comb lines), which is now entirely determined by the frequency noise and spectral separation between the two pinned teeth regardless of whether the actuation of the thermo-refractive index by the main and reference pump is coherent. This can be  demonstrated by leaving the main and reference pump lasers to be free-running. In the fully free-running lasers case, the two pinned teeth correspond to the main and reference pumps, as previously demonstrated~\cite{MoilleNature2023, WildiAPLPhotonics2023}, thereby yielding optical frequency division (OFD) by a factor $\mu_s^2$, and the two lasers do not show any common coherence nor (anti)correlation of their respective actuation of the thermo-refractive effect onto the microring resonator. One can verify the OFD against the same system when the two lasers becomes correlated by locking the CEO through the main pump (i.e., self-referencing), which in this case sees its frequency variation becoming $\Delta \omega_\mathrm{pmp} = \Delta \omega_\mathrm{ref} \times \nicefrac{\left(M_\mathrm{ref} - \mu_s\right)}{M_\mathrm{ref}}$ with $\Delta\omega_\mathrm{ref}$ the variation of frequency of the reference pump,  and the OFD becomes $M_\mathrm{ref}^2 = \left[\nicefrac{\left(\omega_\mathrm{ref} - \omega_\mathrm{ceo}\right)}{\omega_\mathrm{rep}}\right]^2$. 
Experimentally, we use the same microring resonator as in~\cref{fig:2}c, pumped at \qty{\omega_\mathrm{pmp}/2\pi=285.467836}{\THz} \qty{\pm~3}{\MHz} (\qty{\approx 1050.8}{\nm}) and creating a DW at which we send the reference pump at \qty{\omega_\mathrm{ref}/2\pi=192.559616}{\THz} \qty{\pm~2}{\MHz} (\qty{\approx 1557.4}{\nm}), with both frequencies obtained after measuring the repetition rate and CEO as we describe later.  The uncertainty is a one standard deviation value that is dominated by the repetition rate uncertainty. Frequency doubling the reference pump using a periodically-poled lithium niobate waveguide enables $f-2f$ interferometry against the short DW at \qty{385.369148}{\THz} \qty{\pm~4}{\MHz} (\qty{\approx778.4}{\nm}) for CEO detection [\cref{fig:4}a]. The CEO can be efficiently detected with more than \qty{60}{dB} dynamic range as a consequence of the reference pump laser being a comb tooth that provides large power for doubling and the increase in short DW power through the self-balancing effect in KIS~\cite{MoilleNature2023,MoilleFront.Opt.2023}. We use electro-optic modulation to detect \qty{\omega_\mathrm{ceo}/2\pi=-249.91559097}{\GHz} \qty{\pm10}{\Hz} (from electrical spectrum analyzer recording bandwidth) and phase lock it to a \qty{10}{MHz} Rubidium frequency standard, by actuating the main pump frequency through a proportional-integral-derivative (PID) controller [\cref{fig:4}A].
In order to measure the repetition rate \qty{\omega_\mathrm{rep}/2\pi=999.01312}{\GHz} (\qty{\pm10}{\kHz} from its linewidth under free running operation~\cite{MoilleNature2023}), we use a similar electro-optic apparatus that modulates two adjacent Kerr comb teeth. This results in their spectral translation through higher-order modulation sidebands, creating a beat note between a sideband from each Kerr comb tooth that resides within a \qty{50}{\MHz} bandwidth and is detected by a low-noise avalanche photodiode. We proceed to measure $S_\mathrm{rep}$ for three difference cases: single pump~[\cref{fig:4}b(i)], KIS with free-running pumps~[\cref{fig:4}b(ii)], and KIS with a free-running reference and locked CEO~[\cref{fig:4}b(ii)]. In the first case, we observe the $1/f^2$ trend that is a signature of the TRN-limited behavior expected for a single-pumped cavity soliton, as presented above. It is worth pointing out that we use a cooler pump (counterpropagating and cross-polarized relative to the main pump) to thermally stabilize the cavity and provide adiabatic access to the DKS. While a low power cooler (in the 10~mW range) can suppress TRN in microring resonators by about an order of magnitude~\cite{DrakeNat.Photonics2020}, we do not observe such effects. In particular, the simulated $S_\mathrm{rep}^{1p}$ without cooler is orders-of-magnitude lower than our observed noise spectrum~[\cref{fig:4}b(vi)]. We note that for soliton access in our system, the cooler pump must be sufficiently strong, here between \qtyrange{150}{300}{\mW} of on-chip power~\cite{MoilleNat.Commun.2021, moilleTailoringBroadbandKerr2021}, to counteract the thermal shift induced by the 150~mW power main pump and access the DKS state~\cite{ZhangOptica2019, ZhouLightSciAppl2019}. Theoretically, it has been shown that thermal squeezing of integrated cavities works efficiently only at low cooler power, while actually deteriorating the TRN once a critical power threshold is passed~\cite{SunPhys.Rev.A2017}. Therefore, the use of a pump cooler to generate a DKS, although convenient, can greatly degrade its noise metrics when not synchronized. \\
\indent In the KIS regime, $S_\mathrm{rep}^{kis}$ does not follow the TRN profile, as expected from theory, as the frequency noise PSD at \qty{2.5}{\kHz} is brought down from \qty{S_\mathrm{rep}^{1p}\approx4\times10^5}{\Hz^2\per\Hz} to \qty{S_\mathrm{rep}^{kis}\approx 10}{\Hz^2\per\Hz} in the completely free-running KIS case~[\cref{fig:4}b(ii)]. When comparing $S_\mathrm{rep}^{kis}$ and the expected TRN-limited PSD in the single pump case (without the cooler), one can observe that the KIS microcomb has already bypassed the otherwise intrinsic noise limitation of the cavity soliton, beating TRN. We also demonstrate that $S_\mathrm{rep}^{kis}$ is determined by the OFD factor of the KIS operation. In the above case of two free-running pumps, the measured repetition rate noise is well-reproduced by considering the combined pumps' noise and $OFD=\mu_s^2 = 91^2$~[\cref{fig:4}b(ii)]. This demonstrates that the soliton's response to the intra-cavity fluctuation of the microring refractive index is damped regardless of the two lasers' impact on the thermo-refractive shifts of the microring resonances, and is only limited by the phase noise of the two pumps, which in the free-running case is dominant. When the CEO is locked through actuation of the main pump frequency, $S_\mathrm{rep}^{kis}$ is further reduced thanks to the larger $OFD=M_\mathrm{ref}^2 =  193^2$~[\cref{fig:4}b(v)], and is in good agreement with the measured pump frequency noise divided by this larger OFD factor. In this scenario, the reference remains free-running, while TRN is further beaten by a factor of 15 at low frequency. At high frequency, $S_\mathrm{rep}^{kis}$ is limited by the detection floor of the electro-optic apparatus used to measure the $\approx$1~THz repetition rate of the KIS DKS~[\cref{fig:4}b(vii)]. Since our system is limited by the pump(s) frequency noise in both cases, KIS-damped TRN does not constrain the repetition rate performances. We set our reference to be near the KIS window center where $\lambda_\mathrm{ps}\approx-1$. Achieving larger $\lambda_\mathrm{ps}$ values, where KIS internal damping would exceed our noise floor, would place the system too close to the KIS edge, leading to stability issues---an investigation beyond this manuscript's scope.\\

\section*{Conclusion}
In conclusion, we have demonstrated both theoretically and experimentally that Kerr-induced synchronization (KIS) enables, in an all-optical fashion, quenching of internal noise sources in integrated optical frequency combs at a magnitude and with a bandwidth determined by the photon lifetime, thereby making the experimental comb noise metrics mostly dependent on the noise properties of the two pumps, regardless of their shared properties. Thanks to the phase-locking of the cavity soliton onto the reference laser, the comb exhibits a dual-pinning behavior which modifies the so-called elastic-tape model describing the noise propagation across comb teeth. This enables the effective linewidths of the comb teeth to remain within the same order of magnitude as the pump lasers, even for mode separations down to the carrier envelope offset (CEO), in contrast to the single pump soliton case where the comb tooth linewidth can be hundreds of times the pump linewidth at the edges of the comb spectrum, which particularly impacts the linewidth of the CEO. Importantly, the simple frequency noise propagation from tooth to tooth accurately matches the KIS comb and CEO linewidth experiment, while the single pump model diverges from the experiment. This is explained by the importance of thermo-refractive noise (TRN) importance in the single pump case, which in contrast is relatively unimportant due to its quenching in the KIS case. 
To study the impact of the TRN on the microcomb properties, we demonstrate theoretically through a linear stability analysis of the Lugiato-Lefever equation that KIS damps the intracavity noise fluctuations in accordance with the dynamics of attractors in nonlinear damped systems, where noise reduction is determined by the energy exchange rate of the system, here the total loss rate $\kappa$. This holds despite the injection of another laser in the microring resonator, which could potentially raise its temperature and degrade the statistical fluctuations of the cavity refractive index. Hence, regardless of the thermo-optic actuation of the main and the reference pump and their respective correlations, the soliton self-adapts its repetition rate to compensate for the variation of refractive index inside the microcomb to follow the dual-pinning condition. We show that the maximum damping is at the center of the synchronization bandwidth (\textit{i.e,.} where the comb tooth was previously located in a single pump regime). This is confirmed experimentally by measuring the repetition rate noise of the microcomb with free-running lasers, which is found to be entirely determined by the independent pump laser frequency noise and the optical frequency division (OFD) factor from the dual-pinning spacing, while importantly becoming independent of the thermo-refractive noise (TRN) of the cavity. The repetition rate noise is suppressed by more than four orders of magnitude compared to free-running single pumped-DKS, and is about an order of magnitude lower than the minimum TRN that the singly-pumped DKS could exhibit in our system. %
Such demonstration is reproduced when the two lasers becomes correlated, through self-referencing of the microcomb by measuring and locking the CEO by feedback to the main pump. Here, the microcomb repetition rate noise is further decreased (by a factor of 15) according to the larger OFD, and once again is only determined by the frequency noise of the reference laser and not the intracavity noise. %
\\\indent Our work demonstrates that despite the low modal volume of integrated microcombs, which makes them predisposed to large TRN that can greatly hinder their performance, KIS can enable ultra-low-noise performance since the repetition rate noise becomes dependent only on the extrinsic noise properties of the pump lasers and the cavity photon lifetime. This is crucial for octave-spanning combs used in self-referencing, which thus far has only been achieved at high repetition rates (small cavity volumes). Optimization of the photon lifetime can be engineered by leveraging the coupling dispersion of pulley bus-ring couplers~\cite{MoilleOpt.Lett.2019b}, further lowering the noise floor of the KIS microcomb. The external driving noise can be mitigated, as laser locking using on-chip reference cavities has been shown to yield low-noise performance~\cite{LiuOpticaOPTICA2022}. Hence, the noise performance of a laser that is locked to a large cavity can be transferred to the smaller microcomb cavity, with the latter no longer a limitation due to KIS, and thereby providing a pathway for TRN transfer. Our work shows that KIS is a promising approach for the development of low-noise, self-stabilized integrated microcombs with low footprint for SWaP optimization and performance that can potentially become on-par performance with state-of-the-art fiber-based optical frequency combs.

                          
%

\vspace{1em}
\section*{Acknowledgments}
The photonic chips were fabricated in the same fashion as the ones presented in ref.~\cite{MoilleNature2023}, based on a thick Si$_3$N$_4$ process in a commercial foundry~\cite{rahim_open-access_2019}. G.M and K.S acknowledge partial funding support from the AFRL Space Vehicles Directorate and the NIST-on-a-chip program. P.S and C.M. acknowledges support from the Air Force Office of Scientific Research (AFOSR grant FA9550-20-1-0357) and the National Science Fundation (NSF grant ECCS-18-07272). We thank Michael Highman for helping us with experimental instrumentation. We thank Roy Zektzer and Alioune Niang for insightful feedback. The UMBC High Performance Computing Facility (HPCF) was used for some of the simulations presented in this work. G.M thanks T.B.M .

\section*{Author Contributions}
G.M designed the resonator, performed the experimental work, and led the project. P.S developed the theoretical understanding, the analytical framework, and performed the simulation. J.S provided help in the theoretical and experimental noise analysis and understanding. K.S and C.M provided guidance, feedback, and secured funding. G.M, P.S. and K.S. wrote the manuscript, with input from all authors. All the authors contributed and discussed the content of this manuscript.

\section*{Competing Interests}
G.M., C.M. and K.S have submitted a provisional patent application based on aspects of the work presented in this paper.

\section*{Data availability}
The data that supports the plots within this paper and other findings of this study are available from the corresponding authors upon request.

\clearpage
\onecolumngrid
\appendix
\renewcommand{\thesection}{\arabic{section}}
\titleformat{\section}{\large\bfseries}{S.\thesection}{10pt}{}[]
\renewcommand\thefigure{S.\arabic{figure}}  
\renewcommand\theequation{S.\arabic{equation}}
\renewcommand\thetable{S.\Roman{table}}
\renewcommand{\arraystretch}{1.3}
\setcounter{figure}{0} 
\setcounter{equation}{0} 
\renewcommand{\appendixpagename}{%
\centering%
\Large
Supplementary Information: \mytitle
}
\appendixpage

\section{\label{supsec:TRNnoise_analytics}Derivation of analytical repetition rate noise}
Temperature fluctuations in a microring resonator can be described by a Langevin equation,
\begin{equation}
    \label{supeq:lang_temp}
    \frac{d\Delta T}{dt} = -\Gamma_T \Delta T + \zeta_T,
\end{equation}
where $\Gamma_T$ is the thermal dissipation rate and $\zeta_T$ is a white-noise fluctuation source defined by its autocorrelation,
\begin{equation}
    \langle \zeta_T(t) \zeta_T(t + \tau) \rangle = \frac{\Gamma_T k_B T^2}{\rho C V} \delta(\tau).
\end{equation}
with the parameters as defined in the main text after Eq.~(11). Taking the Fourier transform of~\cref{supeq:lang_temp}, we get:
\begin{equation}
    \label{supeq:lang_temp_fourier}
    -i  f \Delta\tilde{T}(\nu) = -\Gamma_T \Delta \tilde{T}(\nu) + W(\nu).
\end{equation}
Therefore,
\begin{align}
    \Delta \tilde{T}(\nu) &= \frac{W(\nu)}{\Gamma_T - i\nu}\\
    S_{\tilde{T}}(\nu) &= \frac{k_B T^2}{\rho C V}\frac{2\Gamma_T}{\Gamma_T^2  + \nu^2}
\end{align}
For small fluctuations in temperature, the change in repetition rate can be written as:
\begin{equation}
    \label{supeq:domega_rep}
    \Delta\omega_\mathrm{rep} = \eta_T \times \Delta T 
\end{equation}
with $\eta_T = \left .\frac{\partial \omega_\mathrm{rep}}{\partial T}\right |_{T_0}$ the variation of the repetition rate with temperature around $T_0$, with more detail provided in~\cref{supsec:TRN_sim}. From~\cref{supeq:domega_rep}, we can determine: 
\begin{equation}
    \Delta \tilde{\omega}_{\mathrm{rep}}(\nu) = \eta_T\frac{W(\nu)}{\Gamma_T - i \nu}
\end{equation}
Hence, when the DKS is under single pump operation (\textit{i.e.} not synchronized to the reference), the power spectral density (PSD) of the frequency noise is given by:
\begin{equation}
    \label{supeq:s_single}
    S_{\omega rep}(\nu) = 2\frac{\eta_T^2k_B T^2}{\rho C V}\frac{\Gamma_T}{\Gamma_T^2  + \nu^2}
\end{equation}

\vspace{1em}
In the Kerr-induced synchronization (KIS) regime, from the linearlization of the LLE presented in the main body of the manuscript, which yields the eigenvalues $\lambda$, one can define another Langevin equation that can be written assuming that fluctuations to the repetition rate damp exponentially,
\begin{equation}
    \frac{\partial\Delta \omega_\mathrm{rep}}{\partial t} = \lambda_{ps}\kappa \Delta \omega_\mathrm{rep} + \eta_T\frac{d\Delta T}{dt}
\end{equation}
where $\kappa = 1/\tau_\mathrm{phot}$ is the total loss rate accounting for the microresonator internal loss (from scattering and absorption) and the coupling loss, and is inversely proportion to the photon lifetime ($\tau_\mathrm{phot}$). Repeating the same operation as in~\cref{supeq:lang_temp_fourier}:
\begin{equation}
    \label{supeq:s_KIS}
   \Delta\tilde{\omega}_\mathrm{rep}(\nu) =\eta_T \frac{i f W(\nu)}{(\Gamma_T - i \nu)(\lambda_{ps}\kappa - i \nu)}
\end{equation}
yielding: 
\begin{equation}
    S_{\omega rep}(\nu) = 2\frac{\eta_T^2k_B T^2}{\rho C V}\frac{\Gamma_T \nu^2}{(\Gamma_T^2  + \nu^2)(\lambda_{ps}^2\kappa^2  + \nu^2)}
\end{equation}

First, it is interesting to note that to minimize $S_{\omega rep}(\nu)$, one needs to minimize $\lambda_{ps}$, which corresponds well with the LLE simulation occurring at the center of the KIS bandwidth, where $\lambda_{ps}=-1$. Secondly, outside of synchronization we obtained $\lambda_{ps}=0$, which correctly makes~\cref{supeq:s_KIS} equal~\cref{supeq:s_single}.

The physical quantities used in simulations, which match the ring resonator design that yields the microcombs in~Figures 2 and 4 are summarized in~\cref{suptab:parameters}.

\vspace{1em}
\section{\label{supsec:MLLE}Multi-driven Lugiato-Lefever equation details}
Here we detail the normalization used to arrive at the Lugiato-Lefever equation (LLE) in~Eq.~(3) of the main text. We assume that one can readily arrive at the LLE from the nonlinear Schr{\"o}dinger equation with periodic boundary condition at every round trip~\cite{ChemboPhys.Rev.A2013}, leading to: 

\begin{equation}
    \frac{\partial a(\theta, t)}{\partial t} = -\left(\frac{\kappa}{2} + i\delta\omega_\mathrm{pmp} \right)a - i\gamma L|a|^2a + i \frac{D_2}{2}\frac{\partial^2}{\partial \theta ^2 }a +  \sqrt{\kappa_\mathrm{ext}P_\mathrm{p}}
\end{equation}

\noindent with $a$ the intractivity field, $\theta$ the azimuthal coordinate of the resonator moving at the speed of the DKS such that $\theta\mapsto \theta - \omega_\mathrm{rep}t$,  $\delta\omega_\mathrm{pmp}$ the pump detuning relative to the cold resonance frequency, $\kappa = 2\kappa_\mathrm{ext}$ the total loss, which in the critical coupling case equals twice the external (coupling) losses, \qty{\gamma = 3.1}{\per\W\per\m} the effective nonlinear coefficient of the system at the \qty{285.5}{\THz} pump frequency (calculated through the finite-element method (FEM)), \qty{L = 2\pi R = 144.51}{\um} the perimeter of the resonator, and $D_2$ the group velocity dispersion of the system. 

One could extend the previous equation in the case of higher-order dispersion coefficients such that $D_\mathrm{int}(\mu) = \sum_{k>1} \frac{D_k}{k!}\mu^k$, with $\mu$ the azimuthal mode relative to the pump, while noticing that $ \sum_{k>1} (-i)^{k+1} \frac{D_k}{k!}\frac{\partial^k}{\partial \theta ^k}a = \sum_\mu D_\mathrm{int}(\mu) A(\mu, t) \mathrm{exp}(i\mu\theta)$, where $A(\mu) = \mathrm{FT}\left[a(\theta, t)\right]$ is the Fourier transform of the intracavity field from $\theta$ to $\mu$.

In addition, a secondary pump can be introduced into the system following ref~\textcite{TaheriEur.Phys.J.D2017} and leading to: 

\begin{align}
    \frac{\partial a(\theta, t)}{\partial t} =& -\left(\frac{\kappa}{2} + i\delta\omega_\mathrm{pmp} \right)a - i\gamma L|a|^2a + \sum_\mu D_\mathrm{int}(\mu) A(\mu, t) \mathrm{e}^{i\mu\theta} \nonumber\\
    +&  \sqrt{\kappa_\mathrm{ext}P_\mathrm{p}} + \sqrt{\kappa_\mathrm{ext}P_\mathrm{ref}}\mathrm{exp}\Bigl[i\bigl(\delta\omega_\mathrm{ref} - \delta\omega_\mathrm{pmp} + D_\mathrm{int}(\mu_s) \bigr)t - i\mu_s \theta\Bigr]
\end{align}

\noindent with $\delta\omega_\mathrm{ref}$ the detuning of the reference relative to its cold cavity, which is offset from the fixed frequency grid spaced by $\omega_\mathrm{rep}$ by $D_\mathrm{int}(\mu_s)$, with $\mu_s$ the mode at which the reference is injected in the cavity and where the synchronization will happen, and $P_\mathrm{ref}$ the reference power in the waveguide. 

Using the change of variable $t \mapsto \frac{\kappa}{2}t$ we obtain~Eq.~(3) of the main text assuming the following normalization in~\cref{suptab:LLEnormparam}: 
\begin{table}[htbp]
    \centering
    \caption{\label{suptab:LLEnormparam}Normalized parameters in~Eq.~(3) of the main text}
    
    \begin{tabularx}{0.5\linewidth}{Y Y}
        \toprule
            Parameter  & Normalization\\
        \midrule
        $\psi$  &  $\sqrt\frac{2 \gamma L}{\kappa}a^*$   \\ 
        $\alpha_\mathrm{p}$ & -$\frac{2}{\kappa} \delta\omega_\mathrm{pmp}$\\
        $\alpha_\mathrm{ref}$ & -$\frac{2}{\kappa} \delta\omega_\mathrm{ref}$ \\
        $\mathcal{D}(\mu)$ & $-\frac{2 D_\mathrm{int}(\mu)}{\kappa}$\\
        $F_\mathrm{pmp}$ & $ \sqrt{\frac{2\kappa_\mathrm{ext}\gamma L }{\kappa}P_\mathrm{p}}$\\
        $F_\mathrm{ref}$ & $\sqrt{\frac{2\kappa_\mathrm{ext}\gamma L }{\kappa}P_\mathrm{ref}}$\\
        \bottomrule
    \end{tabularx}
\end{table}

\vspace{1em}
\section{\label{supsec:TRN_sim}Details on the simulation of the thermo-refractive noise}

Here, we describe the simulation details used to calculate the repetition-rate noise due to TRN using the LLE. The intracavity temperature is modeled as a stochastic variable with a white Gaussian driving force and subject to fluctuation dissipation. This yields the Langevin equation:

\begin{equation}
    \frac{d T}{dt} = -\Gamma_T \Delta T + \zeta_T,
\end{equation}
where $\Gamma_T$ is the thermal dissipation rate, and the temperature driving term $\zeta_T$ is defined by the autocorrelation function, 

\begin{equation}
    \langle \zeta_T(t) \zeta_T(t + \tau) \rangle = \frac{2\Gamma_T k_B T^2}{\rho C V}.
\end{equation}
Fluctuations in the intracavity temperature lead to fluctuations in the pump detunings,

\begin{align}
       \alpha_\mathrm{pmp}(T) &= -\frac{2}{\kappa} \left[\omega_{\mu_0}(T) - \omega_\mathrm{pmp}\right], \\ 
       \alpha_\mathrm{ref}(T) &= -\frac{2}{\kappa} \left[\omega_{\mu_s}(T) - \omega_s\right],
\end{align}
and in the dispersion relation,

\begin{equation}
    \omega_\mu (T) = \omega_\mu(T_0) - (T - T_0)\times \eta_\nu \frac{\omega_\mu^2 (T_0) L_{\text{RT}}}{(\mu + m)c},
\end{equation}
where $\eta_\nu$ is the material thermo-optic coefficient, $L_{\text{RT}}$ is the cavity round-trip length, $m$ is the mode number of the pumped mode, and $c$ is the speed of light in vacuum. The temperature dependent detunings and dispersion relation can be combined into the following expression for the effective integrated dispersion at temperature $T$,

\begin{equation}
    \mathcal{D}(\mu,T) = -\frac{2}{\kappa} \left[\omega_\mu(T) - \omega_{\mu_0}(T_0) - \mu D_1\right],
\end{equation}
where $D_1 / 2\pi$ is the microresonator free spectral range. This leads to the temperature dependent LLE being written as,
\begin{align}
    \label{eq:MLLE_TRN}
    \frac{\partial \psi}{\partial t} &= -\left[1 + i\alpha_\mathrm{pmp}(T)\right]\psi + i\sum_{\mu}\mathcal{D}(\mu,T)\tilde{\Psi} \mathrm{e}^{i\mu \theta} + i|\psi|^2\psi \\
    &+ F_\mathrm{pmp} + F_\mathrm{ref} \exp\Bigl[ i\Bigl(\alpha_\mathrm{ref}(T) - \alpha_\mathrm{pmp}(T) + \mathcal{D}(\mu_s,T)\Bigr)t - i\mu_s\theta\Bigr]\nonumber
\end{align}
The change in repetition rate $\Delta\omega_{\text{rep}}$ due to change in temperature can be calculated directly from the temperature-dependent LLE by tracking the change in the soliton position over time

\begin{equation}
    \Delta \omega_\mathrm{rep}[i] = \frac{\theta_{\text{max}}[i] - \theta_{\text{max}}[i-1]}{\Delta t \times R},
\end{equation}
where $[i]$ denotes the $i^{\text{th}}$ time step. Monte-Carlo simulations were performed to calculate the PSD. The $\Delta \omega_\mathrm{rep}$ data was recorded with a sampling frequency of \qty{10^7}{\Hz} and a frequency resolution of \qty{62.3}{\Hz}. The PSD was calculated by taking the discrete Fourier transform as $|\Delta\widetilde{\omega}_{\text{rep}}(f)|^2$. To obtain a good estimate, 144 parallel simulations were performed at the UMBC HPCF and an averaged PSD was obtained. The quantity $\eta_T$ that is used in \cref{supsec:TRNnoise_analytics} was found by calculating the change in repetition rate locally around the mean temperature $T_0$ using the temperature-dependent LLE as,

\begin{equation}
    \eta_T = \left .\frac{\partial \omega_{\mathrm{rep}}}{\partial T}\right |_{T_0} = \frac{\omega_{\mathrm{rep}}(T_0 + \delta T) - \omega_{\mathrm{rep}}(T_0 - \delta T)}{2\delta T},
\end{equation}
for a small $\delta T$. The parameters used for the TRN simulation are summarized in~\cref{suptab:parameters}

\vspace{1em}
\section{\label{supsec:parameters}Summary of physical parameters}
The physical parameters used in this study to investigate the noise of the microcomb under KIS can be summarized in ~\cref{suptab:parameters} below: 

\begin{table}[h]
    \caption{\label{suptab:parameters} Parameter used for the design of interest considered to study the noise of the microcomb}
    \centering
    \begin{tabularx}{0.8\linewidth}{Y Y Y }
        \toprule
            Parameter  & Values & Method\\
        \midrule
        $k_B$ & \qty{1.38\times 10^{-23}}{\joule\per\kelvin} &  physical constant\\
        $R$ & \qty{23}{\um} &  design\\
        $H$ & \qty{670}{\nm} &  design\\
        $RW$ & \qty{850}{\nm} &  design\\
        $\rho$ & \qty{3000}{\kg\per\m\cubed} &  refs~\citenum{HuszankJRadioanalNuclChem2016, SuyuanBaiIEEETrans.Ind.Electron.2009}\\
        $C$ & \qty{700}{\J\per\kg\per\K} &  refs~\citenum{HuszankJRadioanalNuclChem2016, SuyuanBaiIEEETrans.Ind.Electron.2009}\\
        $V$ & \qty{80}{\um\cubed} &  FEM simulations\\
        $\Gamma_T$ &  \qty{{1\times10^5}}{{\Hz}} & FEM simulations  \\
        $\eta_\nu$ & \qty{2.45 \times 10^{-5}}{\per\kelvin} & ref~\textcite{ArbabiOpt.Lett.OL2013}\\
        $\kappa$ & \qty{{\approx 280\times10^6}}{{\Hz}} &  Measurements\\
        $\eta_T$ & \qty{42 \times 10^6}{\Hz\per\K} &LLE simulations\\
        \bottomrule
    \end{tabularx}
\end{table}

\vspace{1em}
\section{\label{supmat:setup}Experimental setup}
\begin{figure}[H]
    \centering    
    \includegraphics[width=\textwidth]{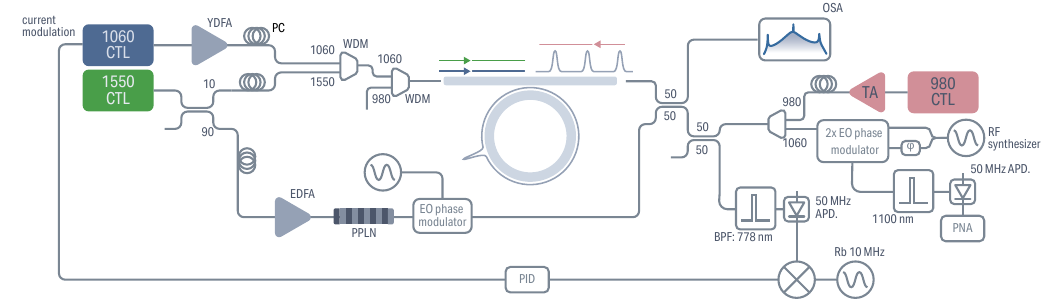}
    \caption{\label{figsup:setup}%
    Experimental setup for the comb generation, repetition rate noise and frequency noise analysis, and carrier envelope offset detection and locking. CTL: continuously tunable laser, YDFA: Ytterbium doped fiber amplifier, EDFA: erbium doped fiber amplifier, WDM: wavelength demultiplexer, PC: polarization controller, TA: taper amplifier, OSA: optical spectrum analyzer, EO: electro-optics, APD: avalanche photodiode, PNA: phase noise analyzer, PPLN: periodically polled lithium niobate, PID: proportional-integral-derivative controller, BPF: band-pass filter.}
\end{figure}
The experimental setup is depicted in \cref{figsup:setup}. We use a \qty{283.5}{THz} main pump provided by a continuously tunable laser (CTL) that is amplified using a ytterbium-doped fiber amplifier (YDFA) to provide \qty{150}{\mW} of on-chip power. This on-chip power accounts for a wavelength-demultiplexer (WDM) and chip insertion losses of about \qty{2}{\dB} per facet when set in the transverse electric (TE) polarization through a polarization controller (PC) to generate the octave-spanning comb spectrum designed for this particular mode. The reference power is provided by another CTL, which is combined with the main pump using a WDM and sent to the chips with the same lensed fiber and also set in TE polarization. The cooler pump, which enables adiabatic tuning of the main pump onto the DKS step, is provided by a third CTL that is amplified using a tapered amplifier (TA) to provide \qty{300}{\mW} of on-chip power at \qty{308.7}{\THz} in a counter-propagating and cross-polarized fashion from the main pump. To do so, we use another pair of 980/1060 WDMs at the input/output to separate the main and cooler pump. In addition a 90/10 coupler is used at the output to process the optical frequency comb without artefacts from the transfer function of the WDM. From there, we use 50~\% of the output signal to probe the comb spectrum with the optical spectrum analyzer (OSA), while the other 50~\% is sent to the electro-optic (EO) comb apparatus, consisting of cascaded electro-optic phase modulators, to spectrally translate two adjacent Kerr comb teeth, to be within a \qty{50}{\MHz} separation from each other. This electrical beat note detected by an avalanche photodiode (APD) is then processed with a phase noise analyzer (PNA). 
For the carrier-envelope offset detection and stabilization, we use another level of 50/50 coupler to tap the high frequency dispersive wave. By sending the frequency doubled reference pump (obtained using a periodically polled lithium niobate (PPLN) crystal) to another EO phase modulator, the large CEO frequency can be measured through interference of the resulting signal with the tapped off high frequency dispersive wave, with a beat note in the \qty{50}{\MHz} band that is detected with another APD. The signal is then locked using a PID controller to feed back onto the main pump frequency and lock the CEO-translated beat to a \qty{10}{\MHz} Rubidium frequency standard.

\section{\label{supsec:linewidth_measurement}Comb linewidth measurement}
\begin{figure}[H]
    \centering    
    \includegraphics[width=\textwidth]{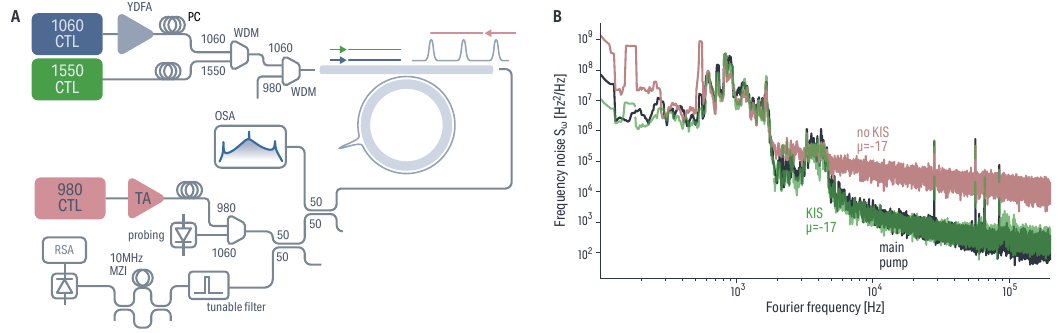}
    \caption{\label{figsup:linewidth}%
    \textbf{(A)} Experimental setup for the comb generation and linewidth analysis of individual comb teeth. CTL: continuously tunable laser, YDFA: Ytterbium doped fiber amplifier, EDFA: erbium doped fiber amplifier, WDM: wavelength demultiplexer, PC: polarization controller, TA: tapered amplifier, OSA: optical spectrum analyzer, RSA: real time spectrum analyser. %
    \textbf{B} Frequency noise power spectral density of the main pump (black) and the comb line at $\mu=-17$ out of synchronizaion (red) and in KIS (green), exhibiting a clear change of the noise profile, from which the effective linewidth is extracted.
    }
\end{figure}
The comb linewidth measurement setup is depicted in \cref{figsup:linewidth}A. The main, reference, and cooler pumps are prepared in the same fashion as in \cref{figsup:setup}. On the output side, however, we use a Gaussian-shape band-pass fiber Bragg grating filter with \qty{0.1}{\nm} full-width half-maximum bandwidth and tunable center wavelength from \qty{980}{\nm} to \qty{1140}{\nm}, which enables us to filter out only one comb line of interest and send it to the \qty{10}{\MHz} unbalanced MZI optical frequency discriminator providing, after processing, the frequency noise of the $\mu^{th}$ tooth [\cref{figsup:linewidth}B]. 
From the frequency noise PSD $S_{\omega}(f)$, we extract the effective linewidth $\Delta\nu_\mathrm{eff}$ such that:
\begin{equation}
    \int_{\Delta \nu_\mathrm{eff}}^{+\infty} \frac{S_{\omega}(\nu)}{\nu^2}\;\mathrm{d}\nu = \frac{1}{\pi}
\end{equation}

\end{document}